\documentclass[%
 reprint,
 amsmath,amssymb,
 aps
]{revtex4-1}

\usepackage{xcolor}
\usepackage{graphicx}
\usepackage{dcolumn}
\usepackage{bm}

\begin{document}

\preprint{APS/123-QED}

\title{Constraining the Yukawa Gravity with Post Newtonian Approximation using S-star Orbits around the Supermassive Black Hole in our Galactic Center}

\author{Yuan Tan}
\author{and Youjun Lu}
\email{luyj@nao.cas.cn}
\affiliation{National Astronomical Observatories, Chinese Academy of Sciences 20A Datun Road, Beijing 100101, China}
\affiliation{School of Astronomy and Space Sciences, University of Chinese Academy of Sciences 19A Yuquan Road, Beijing 100049, China}

\date{\today}

\begin{abstract}

A number of modified gravity theories (e.g., $f(R)$-gravity) lead to a Yukawa-like metric in the weak field limit which can be described by two Yukawa parameters, i.e., the strength $\kappa$ and the length scale $\lambda$. The S-stars, orbiting around the supermassive black hole in the Galactic Center, are unique probes to test these gravity theories in relatively strong gravitational field. The Newtonian Yukawa gravity potential or a simple approximation to the Yukawa metric was usually adopted in previous studies when using the orbital motion of S-stars to constrain such modified gravity theories, which may be not sufficiently accurate considering recent and future high resolution observations. In this paper, we first derive the Post-Newtonian (PN) Yukawa motion equation at the 2PN order, and then investigate the high order effects on the orbital motions by comparison with those from the Newtonian Yukawa gravity potential. We further obtain constraints on $\kappa$ by using the observations on the orbital motions of several S-stars (i.e., S2, S38, and S55). Our results show that the current observations of these stars are compatible with the General Relativity and $\kappa$ can be constrained to $|\kappa|<0.01$ with $95\%$ confidence if $\lambda\in(100,250)$\,AU. We also estimate the possible improvements (about an order of magnitude or more) to the constraints by future higher resolution observations and the inclusion of closer S-stars, such as S4716.

\end{abstract}


\maketitle

\section{Introduction}

The dynamical system in the Galactic Center (GC), consisting of a $4\times 10^6M_\odot$ supermassive black hole (SMBH) and a number of OB type stars (denoted as S-stars) rotating around it, provides a unique laboratory for testing gravity theories \citep[e.g.,][]{2001ApJ...554L..37M, 2009ApJ...703.1743P, 2011ApJ...734L..19A, 2015ApJ...809..127Z, 2016ApJ...827..114Y, 2020ApJ...904..186I, 2021EPJD...75..145J, 2021PhRvD.103h4006Q, 2023arXiv230309284S}. Some S-stars have orbital periods as short as decade(s) or less. For example, the period of the most famous S-star, S2/S0-2 (hereafter S2), is about $16$ years, and its pericenter distance is only $\sim120$\,AU away from Sgr A* \citep{2002Natur.419..694S, 2009ApJ...707L.114G, 2018A&A...615L..15G, 2019A&A...625L..10G}. S38/S0-38 (hereafter S38) and S55/S0-102 (hereafter S55) have periods of $19$\,yr and $12.8$\,yr, respectively, and their pericenter distances are both about $300$\,AU \citep{DATA2, 2017ApJ...845...22P}. Some recently identified S-stars, e.g., S62, S4711, S4714, and S4716, have even shorter periods and smaller pericenter distances \citep{2021A&A...645A.127G, 2021ApJ...918...25P, 2022ApJ...933...49P}. Monitoring the orbital motion of these stars by telescopes (e.g., Keck, Very Large Telescope (VLT)) have already led to tight constraints on the mass of the central SMBH and revealed the relativistic effects induced by it, including the gravitational redshift and periapsis advancement \citep{1996Natur.383..415E, 1997MNRAS.284..576E, 2002MNRAS.331..917E, 2003ApJ...596.1015S, 2009ApJ...692.1075G, 2017ApJ...847..120H, 1998ApJ...509..678G, 2000Natur.407..349G, 2003ApJ...586L.127G, 2005ApJ...620..744G, 2008ApJ...689.1044G, 2003ApJ...594..812G, 2010RvMP...82.3121G, DATA1, GRAVITY2020, 2021arXiv210213000G, 2022A&A...657L..12G}. Future giant telescopes (e.g., Thirty Meter Telescope (TMT) and E-ELT) with sufficiently high astrometric and spectral resolutions are expected to tightly constrain the spin and the metric of the SMBH \citep[e.g.,][]{2008ApJ...674L..25W, 2015ApJ...809..127Z, 2016ApJ...827..114Y, 2017A&A...608A..60G, 2018MNRAS.476.3600W}. 

Various gravity theories have been proposed in order to explain some observation facts, such as the accelerated expansion of  the universe, the rotation curve of galaxies, etc., which are not easy to interpret in the general relativity if not introducing additional components like dark matter and dark energy \citep{2010ARA&A..48..495F, 2005PhR...405..279B, 2014NatPh..10..496S, 2018Galax...6...10D, 2018PhRvD..97j4067D, 2020Univ....6..107D, reviewer..suggest1}. Several theoretical frameworks of such alternative gravity theories arise the Yukawa-like gravitational potential in the weak-field limit. One example is the $f(R)$-gravity, in which the Einstein-Hilbert Lagrangian is extended to be an analytic function of Ricci scalar, $f(R)$ \citep{reviewer..suggest2}. When considering the $\mathcal{O}(2)$-order solution of the field equation in this theory \citep{2012AnP...524..545C}, it leads to a metric with Yukawa-like term. Such a Yukawa-like term can help to explain the flatness in the galactic rotation curves \citep{2018JCAP...08..012D, 2011MNRAS.414.1301C}, and be consistent with observations at smaller scales. A lot of efforts have been made to explore the Yukawa-like term at different scales and environments \citep{1992Natur.356..207F}, including the Galactic Center. 

\cite{fifthforce2017} used the observational data on the orbital motion of S2 and S38 from 1997 to 2013 to constrain the Yukawa potential strength parameter $\kappa$, and they found that $|\kappa|<0.016$ ($95\%$ confidence level) at a length scale of $\lambda=150$\,AU and no discrepancy from the Newtonian gravity at all scales. However, the relativistic effects were not accounted in \cite{fifthforce2017}, as the motion equation they adopted is the Newtonian approximation of the Yukawa-like metric. Since a positive Yukawa term would cause the precession in the same direction as that of the Schwarzschild precession \citep{2018PhRvD..97j4067D}, these two different effects could be mixed, potentially leading to an overestimation of $\kappa$. To get a more precise constraint, \cite{2018PhRvD..97j4068D} and \cite{2021PDU....3300871D} adopted the geodesic equations instead of a modified Newtonian potential. To keep the calculation simple, they adopted an approximation of the metric rather than the original Yukawa-like metric directly deduced from the $f(R)$-gravity. This approximation can deviate from the original metric as large as $20\%$ even when $|\kappa| < 0.1$. As $|\kappa| \ll 0.1$ is needed when explaining binary systems within $f(R)$ gravity \citep{2015IJGMM..1250040D, 2013MNRAS.431..741D}, this approximation can be physically valid. However, considering the recent and future higher resolution observations, delving into the original metric and subsequently deriving a PN approximation could potentially offer a pathway to impose more stringent constraints on $\kappa$ that may be obtained by future observations on the motion of S-stars.

In this paper, we first derive the post-Newtonian (PN) approximation of the motion equation in the Yukawa-like metric resulting from the $f(R)$-gravity in Section~\ref{apprx}. In Section~\ref{model}, we introduce the oribt model for S-stars in details. In Section~\ref{influence}, we illustrate the effects of the PN Yukawa terms on the orbit of an example S-star, i.e., S2. In Section~\ref{data}, we summarize all accessible Keck and VLT observational data on S2, S38, and S55 (from 1992 to 2019) and generate mock data under PN approximation of Schwarzschild metric for future observations by these telescopes as well as the Thirty Meter Telescope (TMT). In Section~\ref{results}, we obtain constraint on the strength parameter $\kappa$ of the Yukawa potential by using all the accessible data and further analyze the improvement on $\kappa$ that may be obtained by future observations. Finally, we summarize our main conclusions in Section~\ref{conclusion}.

\section{Yukawa-like metric and its post-Newtonian approximation}
\label{apprx}

In this Section, we summarize the main steps that lead to the Yukawa-like metric in the $f(R)$-gravity and then derive the second order post-Newtonian approximations for the motion equation in this metric for the first time.  

\subsection{Yukawa-like metric from $f(R)$-gravity}

To clear the logic, we reproduce the deduction of the Yukawa-like metric from \cite{2018PhRvD..97j4068D}. In the $f(R)$-gravity, the natural starting point is to consider a general fourth order gravity action as \citep{2010LRR....13....3D}
\begin{equation}
\mathcal{A}=\frac{c^4}{16\pi G}\int_{}{}\,d^4x\sqrt{-g}f(R)+\mathcal{L}_{\rm m},
\label{con:action}
\end{equation}
where $f(R)$ is an analytic function of the Ricci scalar $R$, $g$ is the determinant of the metric $g_{\mu\nu}$, $c$ is the vacuum speed of light, $G$ is the Newton's constant, and  $\mathcal{L}_{\rm m}$ describes the standard fluid-matter Lagrangian. If $f(R)=R$, it reduces to the classical theory of GR.

From the action in Equation~\eqref{con:action}, the field equation can be derived as
\begin{equation}
f^\prime(R)R_{\mu\nu}-\frac{f(R)}{2}g_{\mu\nu}-f^\prime(R)_{;\mu\nu}+g_{\mu\nu}{\Box}_g f^\prime(R)=\frac{8\pi G}{c^4}T_{\mu\nu}, 
\label{con:field equation}
\end{equation}
where $f^\prime(R)\equiv df(R)/dR$ is the first derivative with respect to the scalar curvature, ${\Box}_g=_{;\sigma}^{\quad;\sigma}$ is the d'Alembertian with covariant derivatives, $T_{\mu\nu}=-2(-g)^{-1/2}\delta(\sqrt{-g}\mathcal{L}_{\rm m})/\delta g^{\mu\nu}$ is the matter energy-momentum tensor. Greek indices run from $0$ to $3$. Taking the trace of Equation~\eqref{con:field equation}, we have
\begin{equation}
3\Box_g f^\prime(R)+f^\prime(R)R-2f(R)=\frac{8\pi G}{c^4}T, 
\label{con:trace equation}
\end{equation}
where $T$ is the trace of $T_{\mu\nu}$.

Following \cite{2012AnP...524..545C}, the solution of field equation can be obtained considering the general spherically symmetric metric
\begin{equation}
\begin{split}
\mathrm{d}s^2&=g_{\sigma\tau}\mathrm{d}x^\sigma \mathrm{d}x^\tau \\
&=-g_{tt}(ct,r)c^2\mathrm{d}t^2+g_{rr}(ct,r)\mathrm{d}r^2+r^2\mathrm{d}\Omega,
\end{split}
\end{equation}
where {$\mathrm{d}\Omega$} is the solid angle. Then, let us consider perturbed metric with respect to a Minkowskian background {$g_{\mu\nu}=\eta_{\mu\nu}+h_{\mu\nu}$}. The perturbed metric can be written as
\begin{equation}
\left\{
\begin{array}{lcl}
g_{tt}(ct,r)\simeq 1+g_{tt}^{(2)}(ct,r)+g_{tt}^{(4)}(ct,r),  \\
g_{rr}(ct,r)\simeq 1+g_{rr}^{(2)}(ct,r), \\
g_{\theta\theta}(ct,r)=r^2,\\
g_{\phi\phi}(ct,r)=r^2\sin^2\theta, 
\end{array}
\right.
\end{equation}
superscripts "(2)" and "(4)" denote the $\mathcal{O}(c^{-2})$ and $\mathcal{O}(c^{-4})$ order contributions in the expansion in powers of $\frac{GM}{rc^2}\sim\mathcal{O}(c^{-2})$. We assume {$f(R)$} is analytic Taylor expandable with respect to a certain value {$R=R_0$}, i.e.,
\begin{equation}
\begin{split}
f(R)&=\sum_{n} \frac{f^n(R_0)}{n!}(R-R_0)^n\\
    & \simeq f_0+f_1R+f_2R^2+f_3R^3+...
\label{con:Taylor expansion}
\end{split}
\end{equation}
Note that at the order {$\mathcal{O}(0)$}, the field equation yield $f_0=0$. Therefore, the solutions at higher orders do not depend on this parameter $f_0$. Consider the $\mathcal{O}(2)$-order approximation, we derive the field equations as
\begin{equation}
\left\{
\begin{array}{lcl}
f_1rR^{(2)}-2f_1g_{tt,r}^{(2)}+8f_2R_{,r}^{(2)}-f_1rg_{tt,rr}^{(2)}+4f_2rR^{(2)}=0,\\
f_1rR^{(2)}-2f_1g_{rr,r}^{(2)}+8f_2R_{,r}^{(2)}-f_1rg_{tt,rr}^{(2)}=0,\\
  2f_1g_{rr}^{(2)}-r \\
\left[f_1rR^{(2)}-f_1g_{tt,r}^{(2)}-f_1g_{rr,r}^{(2)}+4f_2R_{,r}^{(2)}+4f_2rR_{,rr}^{(2)}\right]=0, \\
f_1rR^{(2)}+6f_2\left[2R_{,r}^{(2)}+rR_{,rr}^{(2)}\right]=0, \\
2g_{rr}^{(2)}+r\left[2g_{tt,r}^{(2)}-rR^{(2)}+2g_{rr,r}^{(2)}+rg_{tt,rr}^{(2)}\right]=0.
\end{array}
\right.
\end{equation}

Finally, we obtain the general solution
\begin{equation}
\left\{
\begin{array}{lcl}
g_{tt}^{(2)}=\delta_0-\frac{Y}{f_1r}+\frac{\delta_1(t)\lambda^2e^{-r/ \lambda}}{3 r}+\frac{\delta_2(t)\lambda^3e^{r/ \lambda}}{6r},\\
g_{rr}^{(2)}=\frac{Y}{f_1r}+\frac{\delta_1(t)\lambda^2(1+r/\lambda)e^{-r/ \lambda}}{3r}+\frac{\delta_2(t)\lambda^3(1-r/ \lambda)e^{r/\lambda}}{6r},\\
R^{(2)}=\frac{\delta_1(t)e^{-r/\lambda}}{r}+\frac{\delta_2(t)\lambda^{r/ \lambda}}{2r}. 
\label{con:general_solution}
\end{array}
\right.
\end{equation}
where $\lambda=\sqrt{-\frac{6f_2}{f_1}}$, $f_1$ and $f_2$ are the expansion coefficients, $\delta_0$ and $Y$ are arbitrary integration constants, and $\delta_1(t)$ and $\delta_2(t)$ are two completely arbitrary functions of time with the dimensions of {$\textit{length}^{-1}$} and $\textit{length}^{-2}$, respectively, which can be fixed to constant values,  since the differential equations contain only spatial derivatives. Note that the standard Schwarzschild solution should be recovered in the case of a point-like source with mass $M$ at infinity considering $\lim\limits_{r \to \infty}f(R)=R$, the exponentially growing part in Equation~\eqref{con:general_solution} must be then discarded, which means $\delta_2(t)$ needs to be fixed to zero. Since the integration $\delta_0$ represents an unessential additive quantity for the potential in the integration, it can be set to zero. Requiring that the metric must be asymptotically flat and defining $\frac{2GM}{c^2}=\frac{Y}{f_1}$ and $\delta_1=-\frac{6GM\kappa}{\lambda^2c^2}$, the metric becomes
\begin{equation}
\begin{split}
\mathrm{d}s^2=&-\left[1-\frac{2GM}{rc^2}-\frac{2GM\kappa e^{-r/\lambda}}{rc^2}\right]c^2\mathrm{d}t^2\\
&+\left[1+\frac{2GM}{rc^2}-\frac{2GM\kappa (1+r/\lambda)e^{-r/\lambda}}{rc^2}\right]\mathrm{d}r^2+r^2\mathrm{d}\Omega^2    
\end{split}
\end{equation}

\subsection{The post-Newtonian approximation of Yukawa gravitational potential}

After some manipulations, the Yukawa-like solution can be written as:
\begin{equation}
ds^2=-[1+\Phi(r)]c^2dt^2+[1-\Psi(r)]dr^2+r^2d\Omega,
\label{metric}
\end{equation}
where the two potentials $\Phi(r)$ and $\Psi(r)$ are given by:
\begin{equation}
\left\{
\begin{array}{lll}
\Phi(r)=-\frac{2GM\left({\kappa}\textit{e}^{-\frac{r}{\lambda}}+1\right)}{rc^2}\\
\Psi(r)=\frac{2GM}{rc^2}\left[{\kappa}\textit{e}^{-\frac{r}{\lambda}}+\frac{{\kappa}r\textit{e}^{-\frac{r}{\lambda}}}{\lambda}-1\right]
\end{array}
\right.
\end{equation}
let $r=\rho\left(1+\frac{GM(1-{\kappa}\textit{e}^{-\frac{\rho}{\lambda}})}{{\rho}c^2}\right)$, so $r$ and $\rho$ are asymptotically equivalent. Then we have:
\begin{equation}
\left\{\begin{array}{l}
r^{2} d \Omega=\left(1+\frac{2 G M\left(1-\kappa e^{-\frac{\rho}{\lambda}}\right)}{\rho c^{2}}\right) \rho^{2} d \Omega+O\left(c^{-4}\right) \rho^{2} d \Omega, \\
{[1-\Psi(r)] d r^{2}=\left(1+\frac{2 G M}{\rho c^{2}}\left(1-\kappa e^{-\frac{\rho}{\lambda}} \right) \right) d \rho^{2}+ O\left(c^{-4}\right) d \rho^{2}},  \\
{[1+\Phi(r)] d t^{2}=\left(1-\frac{2 G M}{\rho c^{2}}\left(1+\kappa e^{-\frac{\rho}{\lambda}}\right)\right.} \\
\left.\quad+\frac{2 G^{2} M^{2}}{\rho^{2} c^{4}}\left[1-\kappa^{2} e^{-\frac{2 \rho}{\lambda}}+\frac{\rho}{\lambda} \kappa e^{-\frac{\rho}{\lambda}}\left(1-\kappa e^{-\frac{\rho}{\lambda}}\right)\right]\right) c^{2} d t^{2} \\
\quad+O\left(c^{-4}\right) d t^{2}.
\end{array}\right.
\end{equation}
With the accuracy of $\mathcal{O}(c^{-4})$ and using the usual Cartesian coordinates, the metric becomes:
\begin{eqnarray}
ds^2&=&\Bigg(1-\frac{2GM}{\rho c^2}(1+{\kappa}\textit{e}^{-\frac{\rho}{\lambda}}) \nonumber+\frac{2G^2M^2}{\rho^2c^4}[1-{\kappa}^2\textit{e}^{-\frac{2\rho}{\lambda}}+ \\
&& +\frac{\rho}{\lambda}{\kappa}\textit{e}^{-\frac{\rho}{\lambda}}(1-{\kappa}\textit{e}^{-\frac{\rho}{\lambda}})] \Bigg) c^2dt^2 \\
&&-\left(1+\frac{2GM}{\rho c^2}(1-{\kappa}\textit{e}^{-\frac{\rho}{\lambda}})\right)d\boldsymbol{x}^2. \nonumber
\label{con:rhometric}    
\end{eqnarray}

Divide Equation~\eqref{con:rhometric} by $c^2dt^2$ and define a value of the Newtonian potential $\varphi_N=-\frac{GM}{\rho}$, then Equation~\eqref{con:rhometric} can be written as:
\begin{equation}
\begin{split}
\frac{1}{c^2}\left(\frac{ds}{dt}\right)^2=&1-\frac{\dot{\boldsymbol{x}}^2}{c^2}+\frac{2\varphi_N}{c^2} \left(1+{\kappa}\textit{e}^{-\frac{\rho}{\lambda}}\right)+\frac{2\dot{\boldsymbol{x}}^2\varphi_N}{c^4} \left(1-{\kappa}\textit{e}^{-\frac{\rho}{\lambda}}\right)\\
&+\frac{2\varphi_N^2}{c^4}\left[1-{\kappa}^2\textit{e}^{-\frac{2\rho}{\lambda}}+\frac{\rho}{\lambda}{\kappa}\textit{e}^{-\frac{\rho}{\lambda}}\left(1-{\kappa}\textit{e}^{-\frac{\rho}{\lambda}}\right)\right].
\end{split}
\label{eq:ds2}
\end{equation}

Taking the square root of Equation~\eqref{eq:ds2} with the accuracy of $\mathcal{O}(c^{-6})$, we have
\begin{equation}
\begin{split}
\frac{1}{c}\left(\frac{ds}{dt}\right)
=&1-\frac{\dot{\boldsymbol{x}}^2}{2c^2}+\frac{\varphi_N}{c^2}\left(1+{\kappa}\textit{e}^{-\frac{\rho}{\lambda}}\right)-\frac{\dot{\boldsymbol{x}}^4}{8c^4}+\frac{\dot{\boldsymbol{x}}^2\varphi_N}{2c^4}\left(3-{\kappa}\textit{e}^{-\frac{\rho}{\lambda}}\right)\\
&+\frac{\varphi_N^2}{2c^4}\left[1-2{\kappa}\textit{e}^{-\frac{\rho}{\lambda}}-3{\kappa}^2\textit{e}^{-\frac{2\rho}{\lambda}}+\frac{2\rho}{\lambda}{\kappa}\textit{e}^{-\frac{\rho}{\lambda}}\left(1-{\kappa}\textit{e}^{-\frac{\rho}{\lambda}}\right)\right].
\label{con:rootmetric}    
\end{split}
\end{equation}

Multiplying Equation~\eqref{con:rootmetric} by $-c^2$ and ignoring the leading constant term and those terms with order of  $\mathcal{O}(c^{-6})$ or higher produces a Lagrangian
\begin{equation}
\begin{split}
L=&\frac{\dot{\boldsymbol{x}}^2}{2}\left(1+\frac{\dot{\boldsymbol{x}}^2}{4c^2}-\frac{\varphi_N}{c^2}(3-{\kappa}\textit{e}^{-\frac{\rho}{\lambda}})\right)-\varphi_N \Bigg\{1+{\kappa}\textit{e}^{-\frac{\rho}{\lambda}} \\
&+\frac{\varphi_N}{2c^2}\left[1-2{\kappa}\textit{e}^{-\frac{\rho}{\lambda}}-3{\kappa}^2 \textit{e}^{-\frac{2\rho}{\lambda}}+\frac{2\rho}{\lambda}{\kappa}\textit{e}^{-\frac{\rho}{\lambda}}(1-{\kappa}\textit{e}^{-\frac{\rho}{\lambda}}) \right] \Bigg\}.
\label{con:Lagrangian}
\end{split}
\end{equation}
According to Equation~\eqref{con:Lagrangian} and the Euler-Lagrange equation, we can then derive the post-Newtonian (PN) equation of motion  as
\begin{equation}
\begin{split}
\frac{\textit{d}\boldsymbol{v}}{dt} & =    -\nabla\varphi_N \Big[1+{\kappa}\textit{e}^{-\frac{\rho}{\lambda}}+\frac{4\varphi_N}{c^2}[1-{\kappa}^2\textit{e}^{-\frac{2\rho}{\lambda}}\\
&\quad+\frac{\rho}{2\lambda}{\kappa}\textit{e}^{-\frac{\rho}{\lambda}}(1-{\kappa}\textit{e}^{-\frac{\rho}{\lambda}})]+\frac{\dot{\boldsymbol{x}}^2}{c^2}(1-{\kappa}\textit{e}^{-\frac{\rho}{\lambda}}) \Big]+4(\nabla\varphi_N \cdot \dot{\boldsymbol{x}})\frac{\dot{\boldsymbol{x}}}{c^2}\\
&\quad +\frac{\varphi_N{\kappa}\textit{e}^{-\frac{\rho}{\lambda}}}{\lambda}\frac{\boldsymbol{\rho}}{\rho}\big[1-\frac{\dot{\boldsymbol{x}}^2}{c^2}-\frac{\varphi_N}{c^2}(1-3{\kappa}\textit{e}^{-\frac{\rho}{\lambda}}+\frac{\rho}{\lambda}(1-2{\kappa}\textit{e}^{-\frac{\rho}{\lambda}})\big]\\
&=-\frac{GM}{c^2\rho^3}\Bigg(\bigg[c^2(1+{\kappa}\textit{e}^{-\frac{\rho}{\lambda}})-\frac{4GM}{\rho}[1-{\kappa}^2\textit{e}^{-\frac{2\rho}{\lambda}}\\
&\quad+\frac{\rho}{2\lambda}{\kappa}\textit{e}^{-\frac{\rho}{\lambda}}(1-{\kappa}\textit{e}^{-\frac{\rho}{\lambda}})]+\boldsymbol{v}^{2}(1-{\kappa}\textit{e}^{-\frac{\rho}{\lambda}})\\
&\quad+\frac{\rho}{\lambda}{\kappa}\textit{e}^{-\frac{\rho}{\lambda}}\bigg(c^2-\boldsymbol{v}^{2}-\frac{GM}{\rho}[1-3{\kappa}\textit{e}^{-\frac{\rho}{\lambda}}\\
&\quad+\frac{\rho}{\lambda}(1-2{\kappa}\textit{e}^{-\frac{\rho}{\lambda}})]\bigg) \bigg]\boldsymbol{\rho}-4(\boldsymbol{v}\cdotp \boldsymbol{\rho}) \boldsymbol{v}\Bigg).\\
\label{con:PN}
\end{split}
\end{equation}
Note that if ignoring $\mathcal{O}(c^{-2})$ terms, we recover the Kepler equation of motion with fifth-force correction as
\begin{equation}
\frac{\textit{d}\boldsymbol{v}}{dt} = -\frac{GM}{\rho^3}\left(1+{\kappa}\textit{e}^{-\frac{\rho}{\lambda}}\left(1+\frac{\rho}{\lambda}\right)\right)\boldsymbol{\rho}.
\end{equation}

\subsection{Relative error comparison}

To compare our PN approximation approach with the approximate metric approach proposed by \cite{2018PhRvD..97j4068D}, we calculate the relative errors for both methods. 

For \cite{2018PhRvD..97j4068D}, the approximate metric they used is:
\begin{equation}
ds^2=-[1+\Phi(r)]c^2dt^2+[1-\Phi(r)]dr^2+r^2d\Omega,
\label{approx. metric}
\end{equation}
Thus, their relative error can be defined as:
\begin{equation}
RE=(\Psi-\Phi)/\Phi
\label{RE1}
\end{equation}

For our PN approximation, the error comes from the difference between Equations~\eqref{con:rhometric} and \eqref{metric} and the ignorance of the higher order term $\mathcal{O}(c^{-4})$. The relative error can be defined as
\begin{equation}
RE=\sqrt{ \left(\frac{\Phi_{PN}-\Phi}{\Phi}\right)^2 + \left(\frac{\Psi_{PN}-\Psi}{\Psi}\right)^2 + \left({\frac{GM_{BH}}{rc^2}}\right)^2 },
\label{RE2}
\end{equation}
where 
\begin{equation}
\left\{
\begin{array}{lll}
\Phi_{PN}=-\frac{2GM_{BH}}{rc^2}(1+{\kappa}\textit{e}^{-\frac{r}{\lambda}}) \\
\quad\quad\quad+\frac{2G^2M^2_{BH}}{r^2c^4}[1-{\kappa}^2\textit{e}^{-\frac{2r}{\lambda}}+\frac{r}{\lambda}{\kappa}\textit{e}^{-\frac{r}{\lambda}}(1-{\kappa}\textit{e}^{-\frac{r}{\lambda}})], \\
\Psi_{PN}=-\frac{2GM_{BH}}{rc^2}(1-{\kappa}\textit{e}^{-\frac{r}{\lambda}}).
\end{array}
\right.
\end{equation}
Note that, for S-stars, we have $\frac{GM_{\kappa}}{rc^2} \sim \frac{\boldsymbol{v}^{2}}{c^2} \sim 10^{-4}-10^{-3}$. To estimate the amplitude of these relative errors, we set $\frac{GM_{BH}}{rc^2} = 10^{-3} $ in Equation~\eqref{RE2}.

\begin{figure}
\centering
\includegraphics[width=1.2\columnwidth]{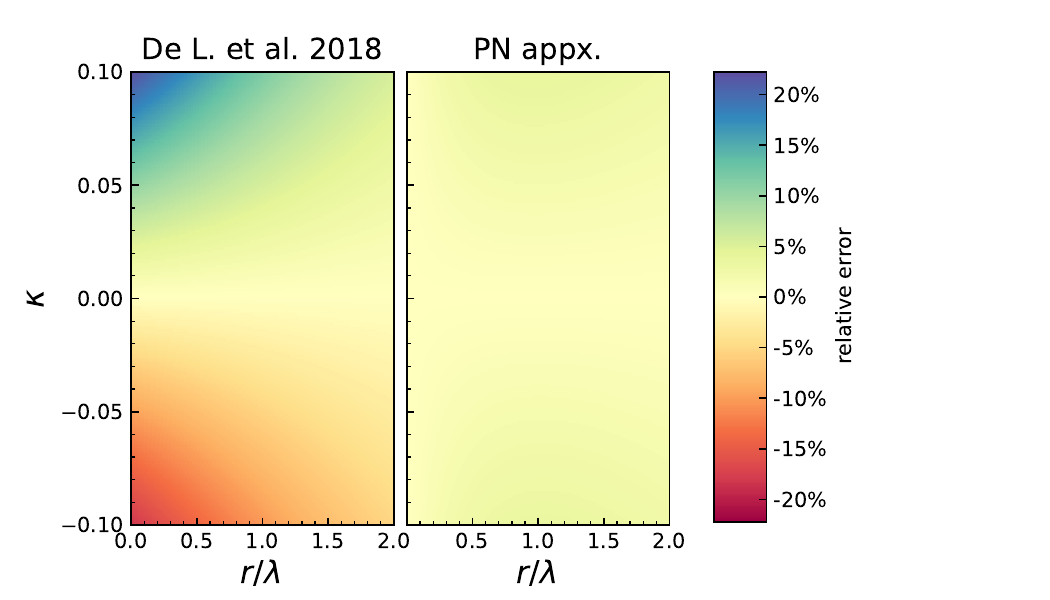}
\caption{
Relative errors of the approximate metric approach \citep{2018PhRvD..97j4068D} and our PN approximation approach as a function of the parameters of the strength $\kappa$ and the scale length ratio $r/\lambda$ of the Yukawa-metric in Equation~\eqref{metric}, calculated using Equation~\eqref{RE1} and Equation~\eqref{RE2}. We set {$\frac{GM_{BH}}{rc^2} = 10^{-3} $} as we are considering S-stars system.
}
\label{fig:compare accuracy}
\end{figure}

We calculate the relative errors for both methods with $\kappa$ in the range from $-0.1$ to $0.1$, and $r/\lambda$ from $0$ to $2$. As shown in Figure~\ref{fig:compare accuracy}, for the approximate metric approach, the relative errors can be as big as $20\%$, while the relative errors of our PN approximation method stay less than $4\%$.

\section{Orbital model}
\label{model}

The orbital model includes the relativistic Doppler effect, gravitational redshift, Rømer delay, and the perturbation due to an extended mass. Below, we describe each of these in details.

\subsection{Relativistic Doppler effect}

As the S-stars' velocities can reach to a few percent of the speed of light, the relativistic Doppler effect must be taken into account. The frequency change due to the relativistic Doppler effect is
\begin{equation}
\omega_0=\omega \frac{1-\frac{v}{c} \cos \alpha}{\sqrt{1-\frac{v^{2}}{c^{2}}}}=\omega \frac{1+\mathrm{RV}/ c}{\sqrt{1-\frac{v^{2}}{c^{2}}}} \approx \omega\left(1+\frac{\mathrm{RV}}{c}+\frac{v^{2}}{2 c^{2}}\right),
\end{equation}
where $\omega_0$ and $\omega$ are the circular frequency of a photon at its emission time and the time when it is received by an observer, respectively,  $\vec{v}$  is the velocity of the source relative to the observer, $\alpha$  is the angle between the velocity vector  $\vec{v}$  and the direction from source to the observer, and $\mathrm{RV}$ is the radial velocity of the source.

\subsection{Gravitational redshift}

S-stars are so close to the SMBH at the Galactic center that the gravitational redshift effect is substantial. The gravitational redshift is given by 
\begin{equation}
\omega_0=\omega \sqrt{-g^{00}} =\omega\frac{1}{\sqrt{-g_{00}}}.
\end{equation}
Since the metric considering in this paper is the Yukawa-like metric, where $g_{00}= -[1+\Phi(r)]$, thus the change of the frequency is
\begin{equation}
\omega_{0}=\omega\frac{1}{\sqrt{1-\frac{2GM({\kappa}\textit{e}^{-\frac{r}{\lambda}}+1)}{rc^2}}} \approx \omega \left( 1+\frac{GM({\kappa}\textit{e}^{-\frac{r}{\lambda}}+1)}{rc^2}\right).
\end{equation}
Taking both the Doppler effect and the gravitational redshift effect into consideration, we can obtain the observed radial velocity as
\begin{equation}
\frac{\mathrm{RV_{obs}}}{c} = \frac{\mathrm{RV}}{c}+\frac{v^{2}}{2 c^{2}}+\frac{GM({\kappa}\textit{e}^{-\frac{r}{\lambda}}+1)}{rc^2}.
\end{equation}
Note that the gravitational redshift effect is roughly on the same order as the transverse Doppler effect as $GM/rc^2 \sim v^2$.

\subsection{Rømer delay}

Since the speed of light is finite, the light travel time from the star to the observer changes during the period, which is known as Rømer delay effect. \cite{2006ApJ...639L..21Z} pointed out that this effect also contributes to $\mathcal{O}(c^{-2})$ order terms, so it is crucial to include it. Following \cite{2018A&A...615L..15G}, we approximate this delay by
\begin{equation}
\begin{split}
t_{\rm e}&=t_{\rm o}-\frac{x(t_{\rm e})}{c}\\
&\approx t_{\rm o}-\frac{x(t_{\rm o})}{c}\left(1-\frac{RV(t_{\rm o})}{c}\right), \\
\end{split}
\end{equation}
where $t_{\rm e}$ denotes the time of emission, $t_{\rm o}$ denotes the time of observation, and $x$ denotes the line-of-sight distance between the source and the observer. For the typical S-star S2, the error of this approximation never exceeds $10$\,seconds, which shows the validity of this approximation.

\subsection{Extended mass}

In order to compare with \cite{fifthforce2017}, we adopt the same extended mass density profile described by a power law as
\begin{equation}
 M_{\rm ext}(<r)=M_{\rm ext}(<r_0)\left(\frac{r}{r_0}\right)^{3-\gamma}.
\end{equation}
The apoapses for S2, S38, S55 are $0.009$\,pc, $0.10$\,pc and $0.007$\,pc, respectively. We fixed the outer radius cutoff {$r_0$} to $0.011$\,pc so that it can enclose all these three stars apoapses. The slope $\gamma$ is set to be $0.5$, considering the fact that the result is not sensitive to its value \citep{fifthforce2017}. 

Considering all the above effects, the model includes following parameters:
\begin{enumerate}
\item six orbital parameters for each star, i.e., the semi-major axis $a$, the eccentricity $e$, the time of the closest approach $T_0$, the argument of periastron $\omega$, the inclination angle $i$, and the longitude of the ascending node $\Omega$;
\item seven SMBH parameters: the mass of the SMBH $M$, the distance to the Galactic Center $R_0$, the position fo the SMBH on the sky $(x_0, y_0)$, and the velocities $(v_{x_0}, v_{y_0}, v_{z_0})$ of the SMBH;
\item two Yukawa parameters: the length scale and strength of the Yukawa gravitational potential $(\lambda, \kappa)$;
\item the total extended mass $M_{\rm ext}$;
\item  the linear difference between the references used by different instruments $(x_k, y_k, v_{x_k}, v_{y_k})$, and the radial velocity offset from NIRC2 $v_{\rm NIRC2}$ \citep{DATA1}.
\end{enumerate}
In total, we have $15+6n$ parameters if we include $n$ S-stars in the orbit modelling. Except for the total extended mass $M_{\rm ext}$, we adopt flat priors for all parameters, and the range for each flat prior is chosen to be wide enough thus the results are not affected by the choice of the priors. Regarding $M_{\rm ext}$, an exponential distribution characterized by a standard deviation of $\sigma_{M_{\rm ext}}=100M_\odot$, whose probability density function is $f(M_{\rm ext})=\frac{1}{\sigma_{M_{\rm ext}}}\exp\left(-{\frac{M_{\rm ext}}{\sigma_{M_{\rm ext}}}}\right)$, is adopted as the prior. This prior is motivated by the observational fact that the total enclosed stellar mass within the S2 orbit may be only a few hundreds of solar mass \citep{2018A&A...609A..27S, GRAVITY2020}. We also note that the mass precession mostly impacts the half orbit around the apocentre \citep{2022A&A...660A..13H}, while the Yukawa effect impacts the whole orbit, allowing for a clear separation of these two effects. Thus the choice of $M_{\rm ext}$ prior would have little influence on the constraints on $\kappa$. 

\begin{figure*}
\centering
\includegraphics[width=\textwidth]{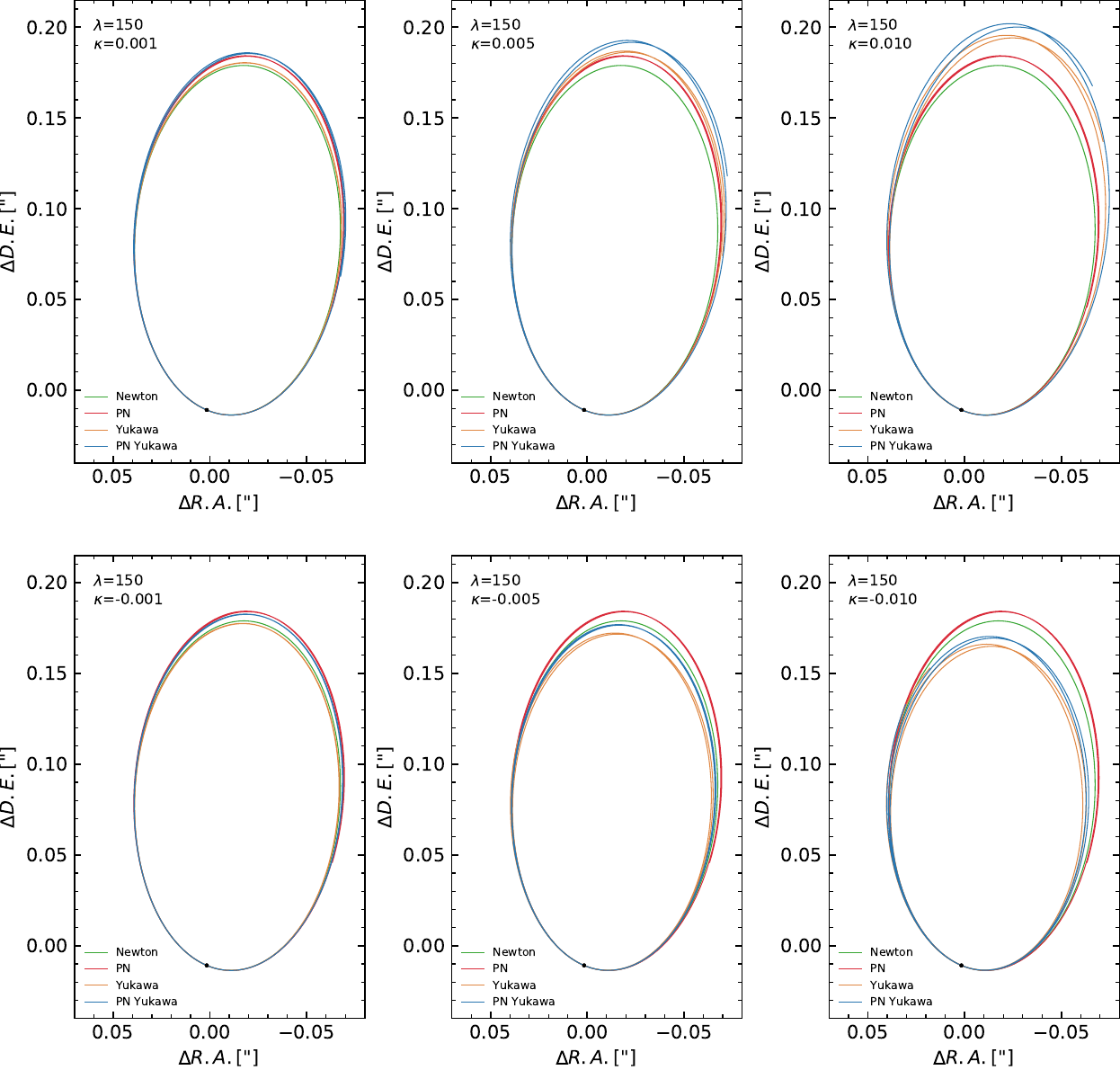}
\caption{
Orbital motion of S2 on the sky calculated by from the Newtonian, PN, Yukawa, and PN Yukawa motion equation. The initial condition for the S2 orbit is adopted by using the fitting result of S2 in \cite{GRAVITY2020} at its latest periapsis passage, i.e., the position and velocity of S2  at $T_0=2018.379$. This starting point is marked by black point. $\lambda$ is fixed to be $150$\,AU, as this is approximately the pericenter distance of S2, around which the Yukawa effect on S2, if any, could be most prominent. The green and red lines show the cases obtained by assuming the Newtonian and PN Schwarzschild motion equation, respectively. The orange and blue lines show the cases obtained by assuming the Newtonian Yukawa and PN approximation Yukawa motion equation, respectively. The absolute value of $\kappa$ are set to $0.001$ (left column), $0.005$ (middle column), and $0.01$ (right column), and the top and bottom panels show the cases with postive and negative $\kappa$, respectively.
}
\label{fig:different_kappa}
\end{figure*}

\begin{figure*}
\centering
\includegraphics[width=\textwidth]{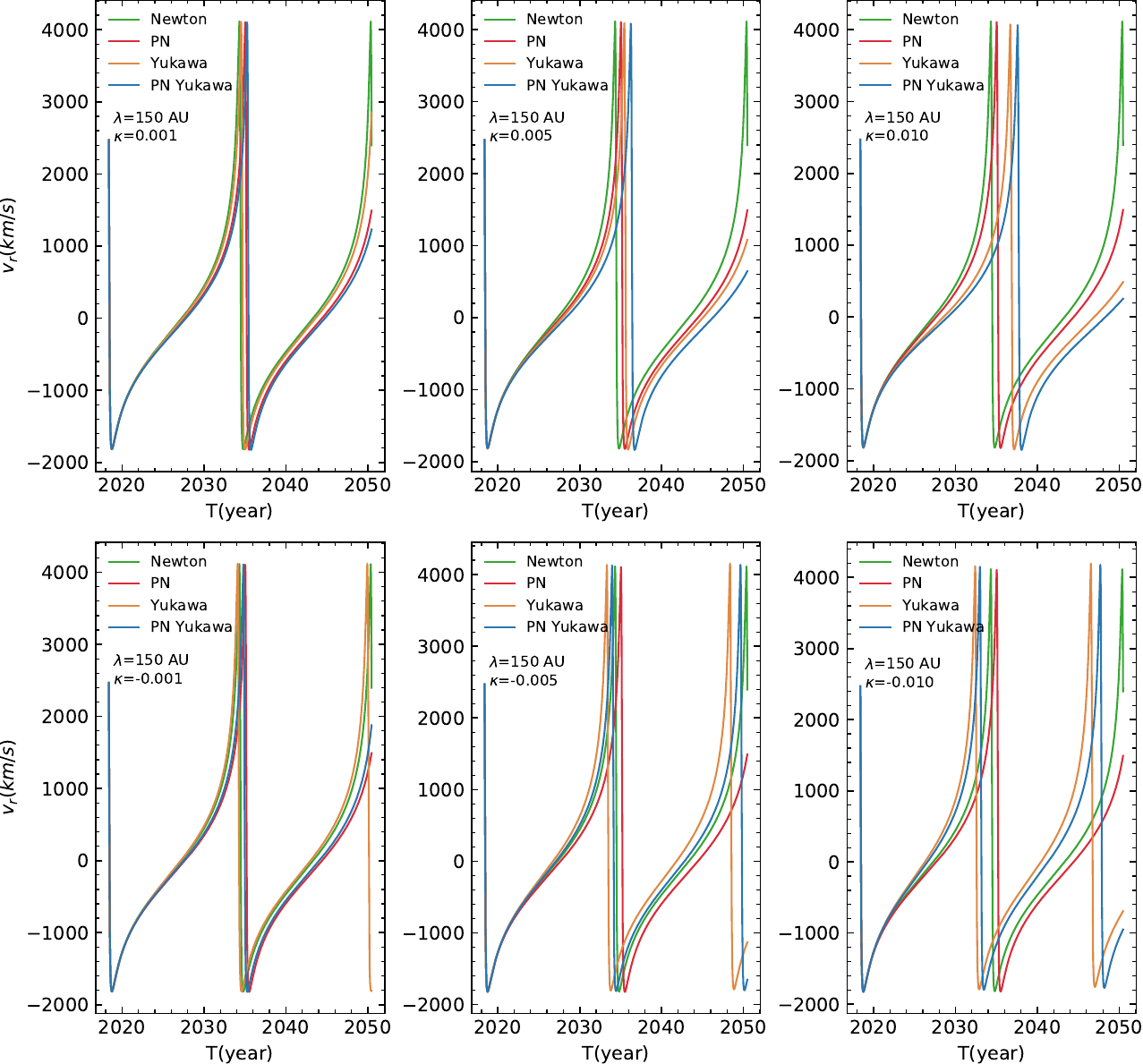}
\caption{
Legends are similar to that for Fig.~\ref{fig:different_kappa}, except for the radial velocity curve of S2.
}
\label{fig:different_kappa_vlsr}
\end{figure*}

\begin{figure*}
\centering
\includegraphics[width=\textwidth]{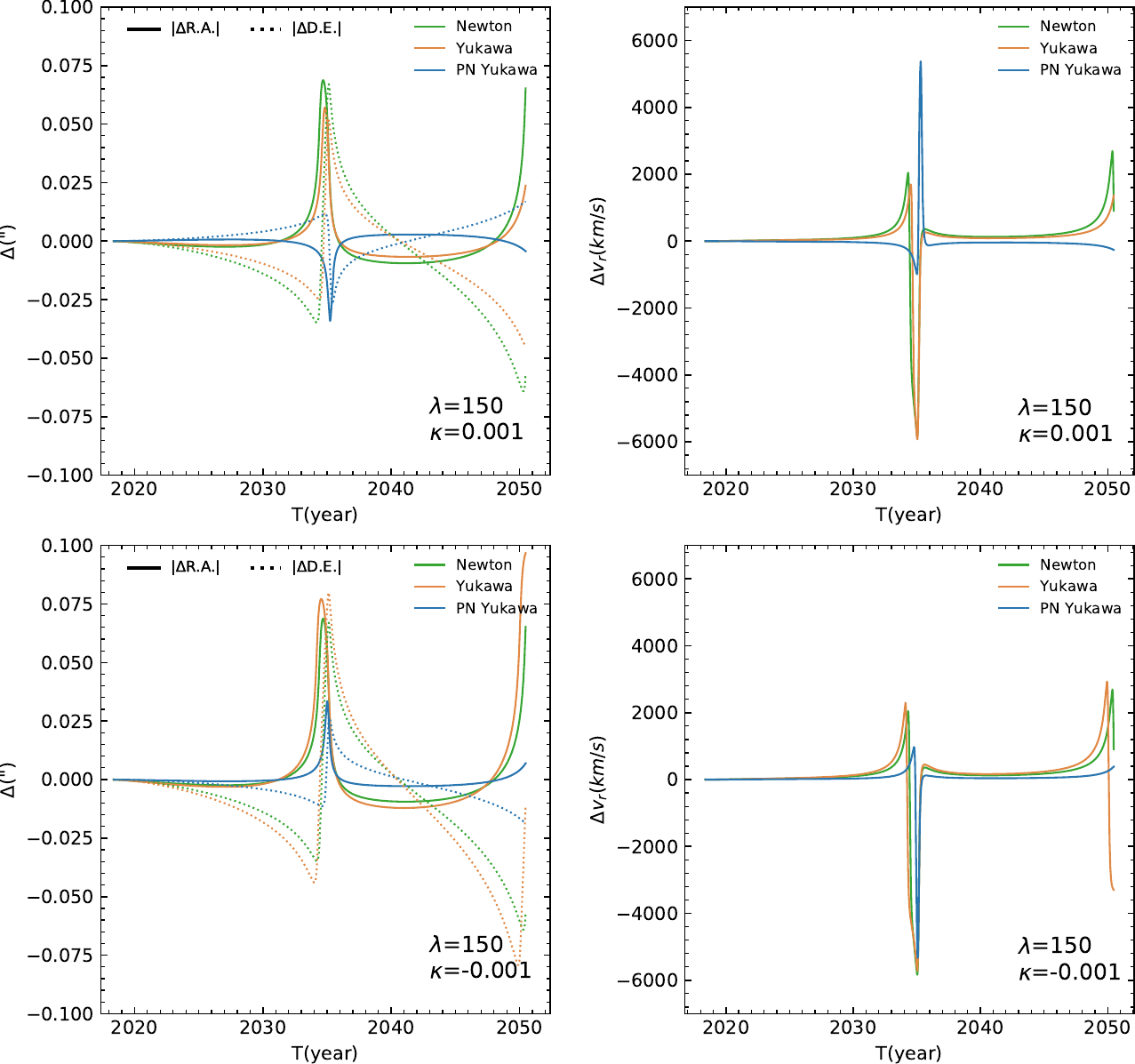}
\caption{
The differences in the apparent position and the radial velocity of S2 between those obtained from the PN motion equation and those obtained from the Newtonian motion equation (green), the Yukawa motion equation (orange), or the PN Yukawa motion equation (blue), respectively. For the cases shown here, $\lambda$ is fixed at $150$\,AU, $\kappa$ is set to be $0.001$ (top panels) or $-0.001$ (bottom panels). In the left panels, the solid lines show the differences in R.A. direction, and the dotted lines show the differences in D.E. direction. 
}

\label{fig:residuals}
\end{figure*}

\section{Influence of the PN Yukawa terms on Orbits}
\label{influence}

We start the integration of the orbital motion equation at $T_0=2018.379$, where S2 is at its periapsis. The initial position and velocity of this star are set by using the semi-major axis and the eccentricity given by \cite{GRAVITY2020} assuming an elliptical orbit in the Newtonian potential. Figure~\ref{fig:different_kappa} shows the expected orbits of S2 on the sky (from $T_0=2018.379$ to $2050$, totally about two S2 orbit periods), calculated from the Newtonian motion equation, PN motion equation, Yukawa metrics, and PN Yukawa  motion equation, respectively, with the same initial conditions.

Note that in this figure, $\lambda$ is fixed to be $150$\,AU, since this is the approximate pericenter distance of S2 and around it the Yukawa effect is the most prominent. As seen from this figure, the Yukawa correction can cause a precession of the S2 orbit, which is different and can be distinguished from that caused by the PN effects. When $\kappa$ is positive, the Yukawa correction  causes the star orbit to precess in the same direction as the Schwarzschild precession and the precession angle per orbit increases with increasing $\kappa$. On the contrary, when $\kappa$ is negative, the Yukawa correction causes the star orbit to precess in the opposite direction to the Schwarzschild precession and the precession angle per orbit increases with increasing value of $|\kappa|$. Furthermore, it is evident that the orbits calculated from the PN approximation and the Newtonian approximation for the Yukawa metrics have significant differences. Besides the Yukawa precession, the former exhibits additional Schwarzschild precession as expected.

In addition, the differences between the Yukawa/PN Yukawa motion equation and the Newtonian/PN Schwarzschild motion equation lead to phase differences. As seen from Figure~\ref{fig:different_kappa_vlsr}, comparing with the velocity curve resulting from the Newtonian and PN motion equation, the Yukawa and PN Yukawa motion equation with a positive $\kappa$ can lead the orbital phase to shift ahead, while those with a negative $\kappa$ can cause orbital phase to shift behind. Figure~\ref{fig:residuals} shows the differences between the apparent position/radial velocity curves of S2 resulting from different motion equations. As seen from this figure, the differences are remarkable especially near the pericenter because of the phase differences resulting from different motion equations, which suggests that the PN Yukawa effect, if any, can be distinguished from other effects.

Note that the situation can be different when $\lambda$ is large. For example, if {$\lambda \gg r$}, the Newtonian approximation equation is reduced to 
\begin{equation}
\frac{\textit{d}\boldsymbol{v}}{dt} \approx -\frac{GM(1+{\kappa})}{r^3}\boldsymbol{r} = -\frac{GM_{\kappa}}{r^3}\boldsymbol{r}, 
\label{big lambda for Newtonian}
\end{equation}
with
$M_{\kappa} = M(1+{\kappa})$.
We can see that it returns back to the Newtonian equation for any value of $\kappa$. Meanwhile, the PN approximation Yukawa equation is reduced to 
\begin{equation}
\begin{split}
\frac{\textit{d}\boldsymbol{v}}{dt} & \approx -\frac{GM_{\kappa}}{r^3}\Bigg(\bigg[1-\frac{4GM_{\kappa}}{c^2r}\frac{(1-{\kappa})}{(1+\kappa)}+\frac{\boldsymbol{v}^{2}}{c^2}\frac{1-{\kappa}}{1+{\kappa}}\bigg]\boldsymbol{r} \\
& -\frac{4(\boldsymbol{v}\cdotp \boldsymbol{r})}{c^2(1+{\kappa})} \boldsymbol{v}\Bigg).
\label{big lambda for PN}
\end{split}
\end{equation}
According to this equation, the PN approximation Yukawa equation reverts to the PN motion equation if $|\kappa| \ll 1$. If $|\kappa| \gg 1$, the Yukawa terms introduces precession in the opposite direction of the Schwarzschild precession.

\section{Observational data and mock data}
\label{data}

\subsection{Observational data}

The real data set used in this work is adopted from \cite{DATA3}, \cite{DATA2}, and \cite{DATA1}, containing $27$ years of Keck and VLT observations of three short period S-stars that have been observed throughout their entire orbits, i.e., S2/S0-2, S38/S0-38, and S55/S0-102. The data set includes two types of data: the apparent positions of the stars on the sky - astrometric observations, and the radial velocities (RVs) of the stars - spectroscopic measurements. The astrometric observations from two different telescopes contain a systematic frame difference which needs linear correction when doing the fitting. All RV measurements are transformed into the local standard of rest frame. Note also that there is a RV offset for data obtained from the NIRC2 spectroscopy with respect to those obtained by other instruments, for which an extra parameter is introduced to describe it \citep{DATA1}.

In total, for S2, we use $45$ and $145$ astrometric observations obtained by Keck and VLT respectively, $115$ spectroscopic measurements of S2 from Keck or VLT; for S38, we use $33$ and $114$ astrometric observations obtained by Keck and VLT respectively, and $6$ spectroscopic measurements of S38 from Keck or VLT; for S55, we use $44$ astrometric observations and $2$ spectroscopic measurements obtained by VLT. 

\subsection{Mock data}
\label{Mock data}

We adopt available observational data by Keck and VLT for these three stars in the past and generate mock observations for S2 in order to investigate the constraints on $\kappa$ and $\lambda$ that may be obtained by current and future observations. 
Note that we do not include the real data from GRAVITY observations in the past (since 2017) when we use the current available observations to do the analysis, as the GRAVITY Collaboration did not release their data yet.\footnote{In principle, one could directly adopt the results on the precession parameter $f_{\rm SP}$ from \citet{GRAVITY2020} to put constraints on $\kappa$ and $\lambda$ similarly as that done in \citet{2021PhRvD.104j1502D}. For simplicity, we do not consider this way to obtain the constraints on $\kappa$ and $\lambda$ in this paper.}
However, for the purpose of checking the improvement in the constraint on $\kappa$ and $\lambda$ by GRAVITY-like future instruments, we generate a mock GRAVITY data set with uncertainties similar to those for the real data (e.g., astrometry precision of $50$\,$\mu$as).
In order to investigate the possible improvements of the constraints on $\kappa$ and $\lambda$ in future by the next generation $30$-m class telescopes with adaptive optics, we further generate mock $32$-year observations of S2 by TMT. In addition, stars that can approach the central SMBH with a distance closer than those of S2, S38, and S55 could be more sensitive to the effect by the Yukawa and PN Yukawa motion equations. For this reason, we also investigate the possible constraints on $\kappa$ and $\lambda$ that may be obtained from future observations of the recently detected S-star, S4716, of which the orbital period is only about $4$ years and the pericenter distance is about $99$\,AU \citep{2022ApJ...933...49P}.

Those mock data are generated by assuming the measurement values following the Gaussian distributions with the expected positions and RVs calculated from PN motion equations as the means and the uncertainties modeled in the following as the variances, respectively.

To describe the astrometric and RV uncertainties, we adopt a trial model, following \cite{futuremodel}, in which the uncertainties are constants for stars with bright magnitudes and follow a power-law for stars with fainter magnitudes, i.e.,
\begin{equation}
\sigma_{\rm astro}(m)=
\begin{cases}
\ A\times10^{\alpha(m-m_1)}, & \text{if\ \ } m>m_1, \\
\ A, & \text{if\ \ }   m\le m_1,
\end{cases}
\end{equation} 
\begin{equation}
\sigma_{\rm RV}(m)=
\begin{cases}
\ B\times10^{\beta(m-m_2)}, & \text{if\ \ } m>m_2, \\
\ B, & \text{if\ \ }  m\le m_2.
\end{cases}
\end{equation}  
In the above equations, the constant $A$ and $B$ are the typical uncertainties that can be reached for bright stars, and the uncertainties that can be reached for fainter stars depend on both the power-law slopes $\alpha$, $\beta$ and the breaking points $m_1$, $m_2$. These parameters are different for different telescopes. Note that $B$, $\beta$, and $m_2$ also depend on the spectral-type of the target star. Since that S-stars are all early-type stars, we only consider the values of these parameters for early-type stars. The values of these parameters adopted in this work are listed in Tables~\ref{con:astro-uncertainty} and \ref{con:rv-uncertainty} (see also \cite{futuremodel}).

According to this simple model, the astrometric and RV uncertainties of the S2 posisition and radial velocity given by mock observations of GRAVITY and Keck/VLT are roughly $\sim 0.05$\,mas and $12$\,km/s, which are similar to the uncertainties in the current observations, as seen in \cite{2023arXiv230304067T}.

With respect to the observation strategy, we assume that the observations would be more frequent around pericenter, and more infrequent elsewhere, followed \cite{2022JCAP...03..007D}. In details, we assume the observations performs:
\begin{itemize}
\item One observation per day in two weeks centered on the pericentre;
\item One observation every two nights in the rest of the month centered on the pericentre;
\item One observation per week in the rest of the two months centered on the pericentre;
\item One observation per month in the rest of the year centered on the  pericentre;
\item Two observations per year in the rest of the years.
\end{itemize}

\begin{table}
\centering
\caption{Astrometric uncertainties for different telescopes.}
\begin{tabular}{|c|c|c|c|} \hline
\multicolumn{4}{|c|}{Astrometric uncertainty} \\ \hline
Telescope     & A({$\mu$}as)     & {$\alpha$}      & {$m_1$}    \\ \hline
GRAVITY (VLTI)           & 50         & 0.2    & 15      \\ \hline
TMT           & 25         & 0.2    & 17      \\ \hline
\end{tabular}
\label{con:astro-uncertainty}
\end{table}    

\begin{table}
\centering
\caption{RV uncertainties for different telescopes.}
\begin{tabular}{|c|c|c|c|}
\hline
\multicolumn{4}{|c|}{Early-type star RV uncertainty} \\ \hline
Telescope  & B(km/s)  & {$\beta$}  & {$m_2$} \\ \hline
Keck/VLTI & 12       & 0.17                  & 14   \\ \hline
TMT        & 3        & 0.17                  & 14   \\ \hline
\end{tabular}
\label{con:rv-uncertainty}
\end{table}

\section{Results}
\label{results}
 
We adopt the MCMC bayesian sampler (the \textit{emcee} python package \citep{emcee}) to perform the fittings to the observations data and the mock data of the orbital motion of the target stars. To make sure the samples are well converged, we require the sample longer than $30$ times of the autocorrelation time per chain and the Gelman-Rubin diagnostic of $R-1 < 0.05$ \citep{2018ApJS..236...11H}. By sampling the posterior probability distribution function of $\kappa$ for a given $\lambda$, we can determine a $95\%$ confidence upper limit on its absolute value.

\subsection{Possible constraints on $\kappa$ for a given range of $\lambda$}

\begin{figure}
\centering
\includegraphics[width=\columnwidth]{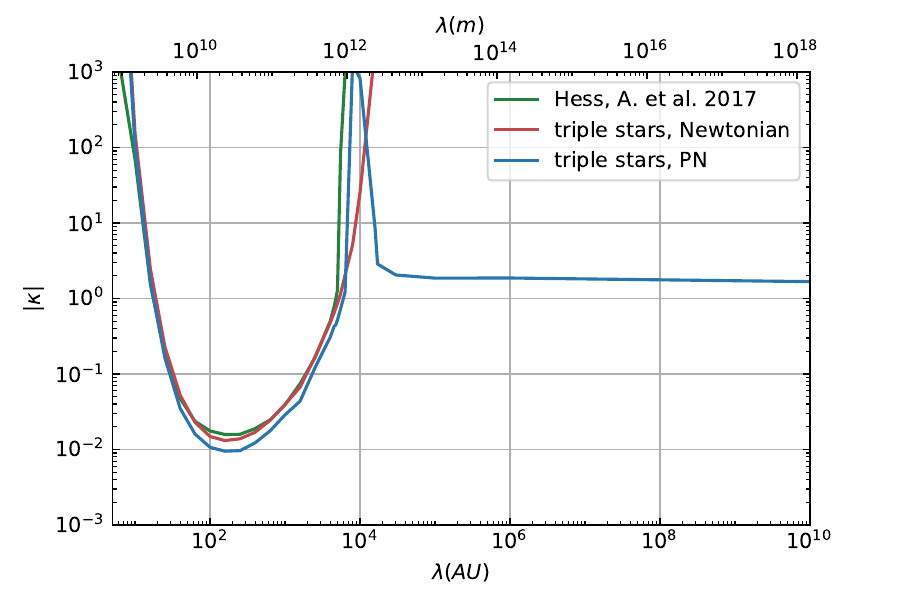}
\caption{A $95$\% confidence upper limit of Yukawa strength $\kappa$ as a function of $\lambda$ constrained by the orbital motion observation of S2, S38, and S55 over a period of $30$ years. The results deduced from the PN (Newtonian) Yuakwa motion equation are shown by the blue (red) line. For comparison, the result obtained by \cite{fifthforce2017} is also shown by the green line.}
\label{fig:fitresult}
\end{figure}

We first obtain constraints on $\kappa$ by using the available observations on the orbital motions of S2, S38, and S55, as shown in Figure~\ref{fig:fitresult}. As seen from this figure, $|\kappa|$ is constrained to be around $0$ within the $68\%$ confidence interval if $\lambda \in (5, 10^{10})$\,AU, indicating no significant deviation from the GR theory. The strongest constraint is around $\lambda\sim150$\,AU, which is dominated by the observations of S2. If $\lambda=150$\,AU, the observational data suggests a $95\%$ confidence upper limit of $|\kappa|<0.009$, if adopting the PN Yukawa motion equations, 
and $|\kappa|$ can be constrained to $|\kappa|<0.01$ with $95\%$ confidence if $\lambda \in (100,250)$\,AU.
This constraint is stronger than that derived by using the traditional Newtonian Yukawa approximation of motion equations \citep{fifthforce2017}. This is partially ascribed to the new data near pericenter and the involvement of S55, but mainly due to the fact that, instead of treat all precessions as the Yukawa precession, the total precession now is devided into the Yukawa precession and the Schwarzschild precession, which is included by the PN terms in the motion equations. Therefore, a smaller $\kappa$ would suffice to generate the same amount of total precession. 

We can also see that the constraining power on $\kappa$ becomes weaker if $\lambda$ is around $10^4$\,AU. At this scale, $\lambda$ is much larger than the pericenter distance of S2 ($\sim120$\,AU), but has similar magnitude as its apocenter distance ($\sim2000$\,AU). Thus the Yukawa effect would perform differently at different stage of the orbit when $\kappa\gg1$. Near the pericenter (apocenter), the Yukawa effect causes a precession in the opposite (same) direction of the Schwarzschild precession.The counteraction between these two precessions results in the weaker constraining power at this $\lambda$ range. 

In addition, Figure~\ref{fig:fitresult} also shows a remarkable difference between the curves obtained by using the Newtonian Yukawa motion equations and those obtained by using the PN approximations of the Yukawa motion equations for large $\lambda$. With $\lambda$ increasing to $\gtrsim 10^4$\,AU, the red curve representing the constraints obtained by using the Newtonian approximation equation goes to infinity, while the one derived from the PN equations maintain a finite value. 
This feature is also found by \cite{2021PhRvD.104j1502D}, in which the orbital precession deduced from the approximation Yukawa metric is adopted to constrain the Yukawa parameters. As clearly shown by Figure~1 in  \cite{2021PhRvD.104j1502D}, for large values of $\lambda$, $|\kappa|$ cannot be constrained at $95\%$ confidence if the precession measurements are excluded (top panel), and it can be constrained if the precession measurements are included (bottom panel).
This difference actually can be understood according to Equations~\eqref{big lambda for Newtonian} and \eqref{big lambda for PN}. The degeneracy between {$GM$} and {$(1+\kappa)$} (see Eq.~\eqref{big lambda for Newtonian}) suggests that $\kappa$ is impossible to be  constrained by the observations if $\lambda$ is large. If adopting the PN approximation to the Yukawa motion equations, however, this degeneracy can be broken (see Eq,~\eqref{big lambda for PN}). Therefore, even for large values of $\lambda$, $|\kappa|$ can still be limited to $<2$ without obviously dependence on $\lambda$.

\begin{figure}
\centering
\includegraphics[width=\columnwidth]{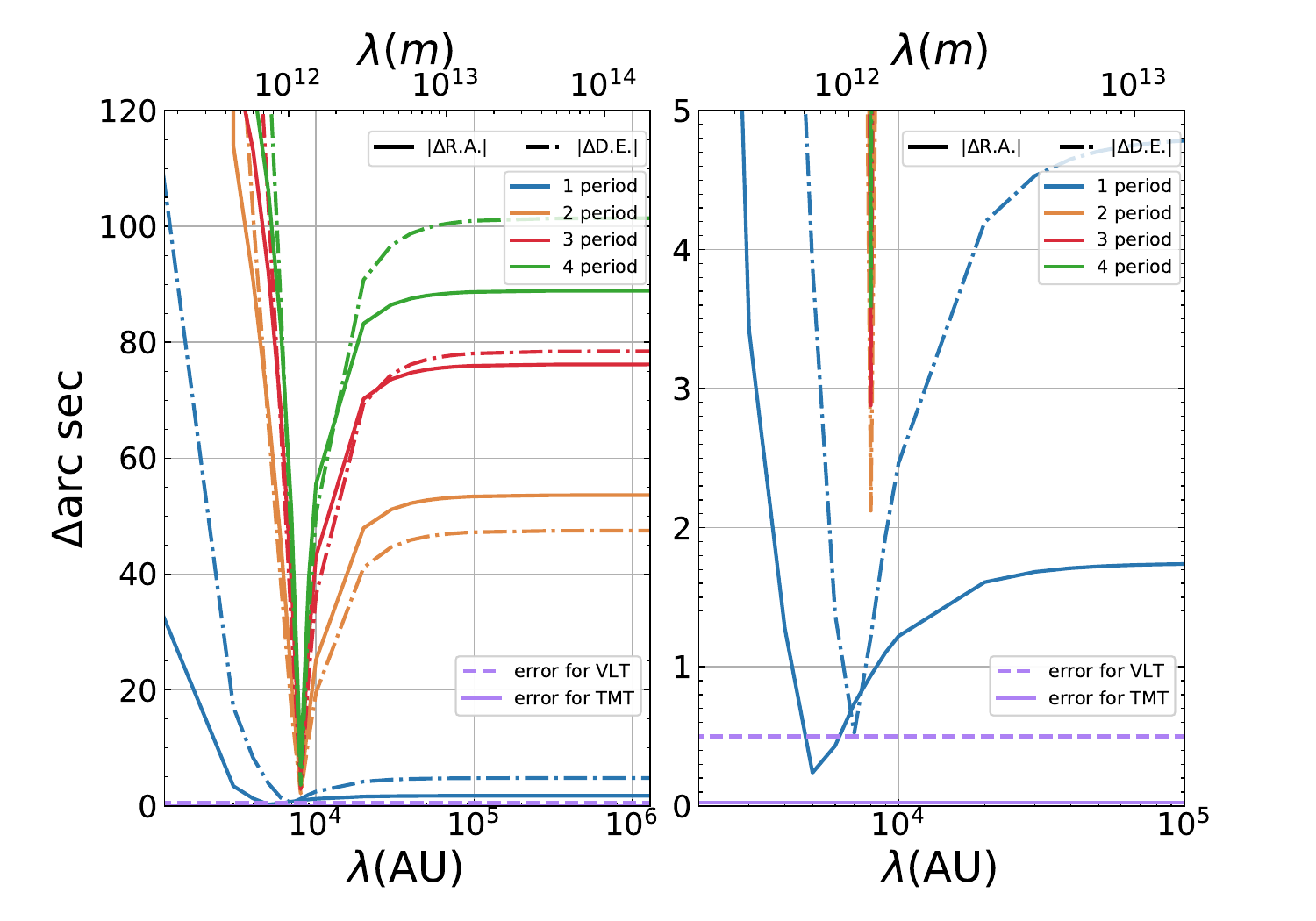}
\caption{
The maximum values of differences between the S2 positions on the sky at any given time resulting from the PN Yukawa motion equations (with $\kappa=1$ but different $\lambda$) and those from the PN motion equations derived from the Schwarzschild metric over an observation period. The right panel shows a zoom-in part of the left panel with $\Delta \text{arc sec} \le 5$. The blue, orange, red, and green lines indicate the maximum differences over the observation period of one, two, three, and four orbits, respectively. The solid lines show the deviations in the R.A. and the dash-dotted lines illustrate the deviations in D.E.. The purple solid and dash lines represent the observation errors for VLT and TMT, respectively. 
}
\label{fig:lambda_vs_Delta}
\end{figure}

To further illustrate the feature of curves shown in Figure~\ref{fig:fitresult}, we calculate the maximum differences between the S2 orbits deduced from the PN Yukawa equation and PN equation for varying $\lambda$ by fixing $\kappa$ at $1$, as shown in Figure~\ref{fig:lambda_vs_Delta}. As seen from this figure, the differences are reduced to a minimum when $\lambda \sim 6 \times 10^3$\,AU, after which it increases with increasing $\lambda$ until no further changes are observed for larger $\lambda$ values. If we only observe for one period, the difference around $\lambda \sim 6 \times 10^3 $\,AU can be substantially smaller than the errors in current observations, which thus makes it impossible to constrain $\kappa$ to be lower than $1$. However, as $\lambda$ increases, the differences increase and exceed the errors in current observations. Therefore, the current observations can put a constraint on $\kappa$ to be lower than $1$ (see also Fig.\ref{fig:fitresult}). In addition, if we extend the observation period substantially or use observations by future telescopes, it is possible to further constrain $\kappa$ to smaller values around $\lambda\sim6 \times 10^3$\,AU.

\subsection{Future improvements on $\kappa$ constraints}

In this section, we further demonstrate the possible constraints on $\kappa$ that may be obtained by the observation of S-star orbits by future telescopes using the mock observations (see Section~\ref{data}). \cite{2023arXiv230304067T} pointed out that the calibration of the reference frames between different instruments can be degenerate with the problem, therefore we only use those high resolution mock data sets obtained by different telescopes separately in this part. 

\begin{figure}
\centering
\includegraphics[width=\columnwidth]{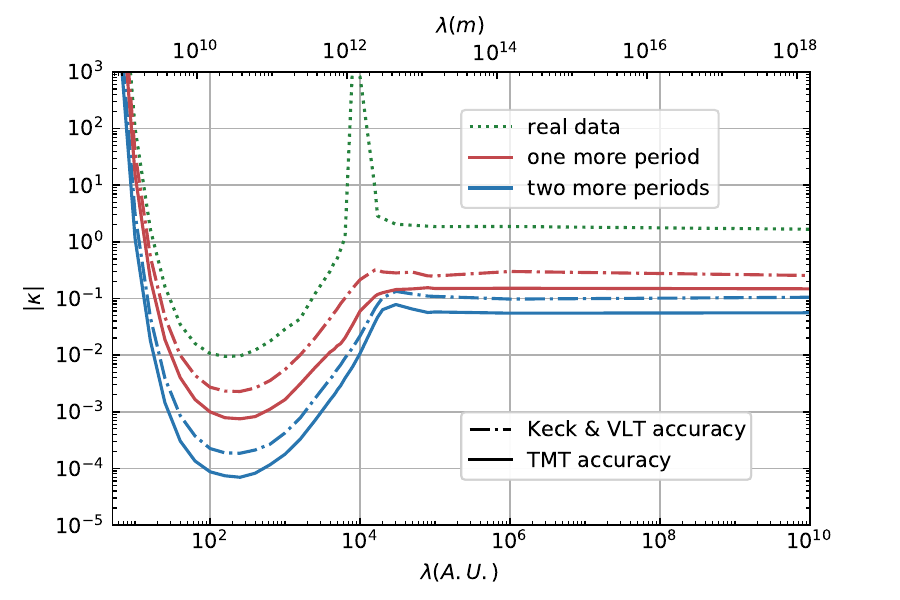}
\caption{
Expected constraints on $|\kappa|$ by future observations on the orbital motion of S2 over a long period. The red and blue lines show the results obtained from the observations by Keck$\&$VLT and TMT, respectively. The dash-dotted and solid lines show the results by using the observations over a period of 16 years and $32$ years, respectively. For comparison, the constraint given by current available observations is also shown in the figure by the green dotted line.
}
\label{fig:futureS2}
\end{figure}

Figure~\ref{fig:futureS2} shows the expected constraints on $|\kappa|$ for different $\lambda$ by using long period observations of the orbital motion of S2 via Keck and VLT or TMT. Each additional period, i.e. $\sim 16$ years, of observation can enhance the constraint by nearly one magnitude. Not surprisingly, the constraint expected from the TMT observations is stronger compared with those from Keck and VLT. With observations over two orbital periods by TMT, $|\kappa|$ can be limited to $<10^{-4}$ around $\lambda\sim150$\,AU, which is more than two orders of magnitude higher that that obtained by current observations. 

\begin{figure}
\centering
\includegraphics[width=\columnwidth]{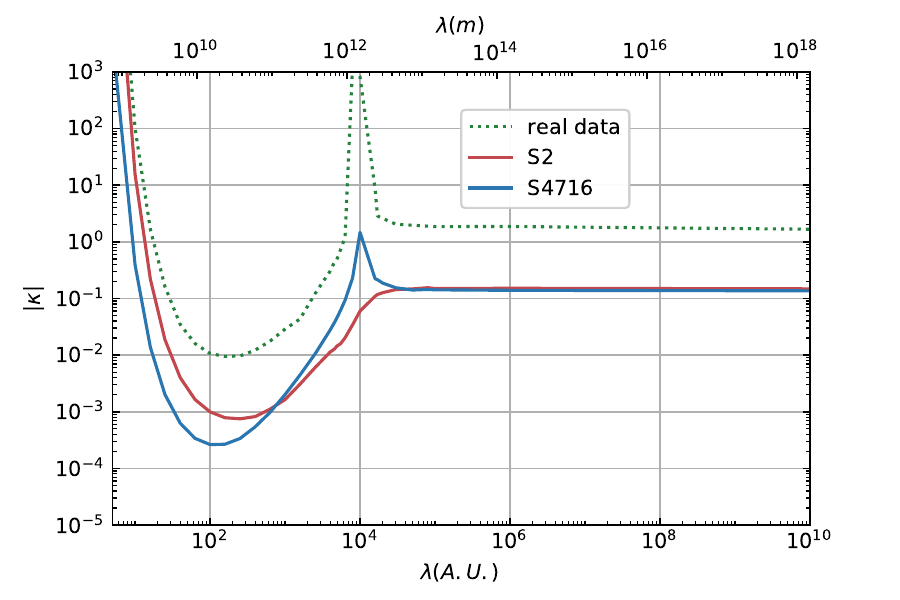}
\caption{
Expected constraints on $|\kappa|$ by future TMT observations on S4716 over a period of $16$ years (blue line). For comparison, the constraint obtained by using the current available data (green dotted line) and by using the TMT observations of S2 over $16$ years (red line) are also shown in the figure.
} 
\label{fig:futureS4716}
\end{figure}

Short period S-stars can reach the central SMBH in a closer distance than S2, which may lead to stronger constraints on $\kappa$. S4716 has a period of only about $4$ years and a pericenter distance of $\sim 99$\,AU, short and smaller than those of S2, which could be a nice observational objective. We generate mock TMT observations for the orbital motion of S4716 over a period of $16$ years and using the same method as done for S2 to investigate the possible constraints that may be obtained from it. As shown in Figure~\ref{fig:futureS4716}, with $16$ years observations of S4716 by TMT, $|\kappa|$ could be constrained to $< 2.6 \times 10^{-4}$, while the same time span of TMT observations for S2 can only lead to $|\kappa| < 8 \times 10^{-4}$. It is clear that the observations of stars like S4716 can lead to a substantial improvement on the constraint of $\kappa$ and the strongest constraint that can be obtained shift to the region with slightly smaller $\lambda$, compared with those by S2. 

\section{Conclusions and discussions}
\label{conclusion}

In this paper, we investigate the constraints on the Yukawa metric that may be obtained by current and future observations on S-stars rotating around the SMBH in the Galactic center. Starting from the $f(R)$-gravity, we first derive the 2PN approximation to the motion equation in the Yukawa metric, characterized by two Yukawa parameters, i.e., the Yukawa strength $\kappa$ and the length scale $\lambda$. This equation can reach to higher order (2PN order) of accuracy compared with the traditional Newtonian approximation of the motion equations in the Yukawa metric. This PN approximation to the motion equation can lead to both the Schwarzschild precession and the Yukawa precession, while the traditional Newtonian approximation to the Yukawa metric only leads to the Yukawa precession.

We further use all the accessible observational data of S2, S38, and S55 to obtain constraint on $\kappa$ by adopting the PN approximation of the Yukawa motion equation via the MCMC method. Since the PN Yukawa motion euqation can separate the Yukawa precession from the Schwarzschild precession, we obtain a stronger constraint on $\kappa$ than that by previous studies \citep{fifthforce2017}. We find that $|\kappa|<0.009$ if $\lambda\sim150$\,AU and $|\kappa|<0.01$ if $\lambda \in (100,250)$\,AU at the $95\%$ confidence level. We emphasize here that the degeneracy between the SMBH mass and the Yukawa strength can be broken by using the PN Yukawa motion equation and the Yukawa strength $\kappa$ can be constrained to be substantially less than $2$ even if $\lambda$ is large. 
This result is consistent with that in \cite{2021PhRvD.104j1502D}.
Furthermore, we demonstrate that the constraints on $\kappa$ can be improved by more than one to two orders of magnitude at $\lambda \sim 100-1000$\,AU by using future Keck$\&$VLT and TMT observations of S2 over one to two orbital periods ($16$ to $32$ years). If using the TMT observations of a star like S4716, with a orbital period substantially smaller than that of S2, one may be able to obtain better constraints on $\kappa$ at slightly smaller $\lambda$ than those by S2 with the same period of observations.

Note that when using upcoming higher resolution data and studying stars with smaller semi-major axis, the higher order  effects of gravity, not considered in the above analysis, could be no longer negligible. For example, the frame-dragging effect caused by the spin of Sgr A* could be detectable by long-time monitoring of S-stars, if the precisions of astrometric and radial velocity measurements can be as high as $({\sigma }_{{\rm{p}}},{\sigma }_{Z})\sim (10\;\mu \mathrm{as},1\;\mathrm{km}\;{{\rm{s}}}^{-1})$ \citep[e.g.,][]{2015ApJ...809..127Z, 2016ApJ...827..114Y}. Furthermore, the quadruple moment, dependent on both the absolute value and the direction of the spin, can also lead to higher order of precessions. In addition, the light could be bent by the central SMBH, which affects the apparent positions of the target star on the sky \citep{2012ApJ...753...56B, 2015ApJ...809..127Z, 2017A&A...608A..60G}. 

The PN Yukawa motion equation may also be tested in other systems. For example, the Solar System is another good laboratory for testing gravity theories. As the precessions of planets are observational determined with high accuracies, it is necessary to adopt the PN Yukawa motion equation when constraining $\kappa$ at the Solar System scale by using these observations. 
Additionally, pulsars can provide another method to test gravity theories. The post-Keplerian parameters of some binary pulsar systems have been measured nowadays \citep{2006Sci...314...97K}. As the discovery and measurement of binary pulsars are continuing, the PN Yukawa motion equation could also be tested in some binary pulsar systems \citep{2022JCAP...11..051D}. 

\begin{acknowledgments}
This work is partly supported by the National Natural Science Foundation of China (grant nos. 11991052, 12273050, 11690024), the Strategic Priority Research Program of the Chinese Academy of Sciences (grant No. XDB0550300), and the National Key Program for Science and Technology Research and Development (grant nos. 2022YFC2205201, 2020YFC2201400).
\end{acknowledgments}

\bibliographystyle{apsrev4-1}
\bibliography{ref}

\end{document}